\documentclass[twocolumn,aps,amsmath,superscriptaddress,amssymb,pre]{revtex4-1}

\usepackage[dvips]{graphicx}
\usepackage{latexsym}
\usepackage{natbib}

\begin{document}

\bibliographystyle{apsrev}

\title{How people interact in evolving online affiliation networks}

\author{Lazaros K. Gallos}
\affiliation{Levich Institute and Physics
    Department, City College of New York, New York, New York 10031,
    USA}

\author{Diego Rybski}
\affiliation{Levich Institute and Physics
    Department, City College of New York, New York, New York 10031,
    USA}
\affiliation{Potsdam Institute for
    Climate Impact Research, 14469 Potsdam, Germany}

\author{Fredrik~Liljeros}
\affiliation{Department of Sociology, Stockholm
    University, S-10691, Stockholm, Sweden}

\author{Shlomo~Havlin}
\affiliation{Minerva Center and Department of Physics,
    Bar-Ilan University, Ramat Gan 52900, Israel}

\author{Hern\'an~A.~Makse}
\affiliation{Levich Institute and Physics
    Department, City College of New York, New York, New York 10031,
    USA}

\date{\today}

\begin{abstract}
  The study of human interactions is of central importance for
  understanding the behavior of individuals, groups and societies.
  Here, we observe the formation and evolution of networks
  by monitoring the addition of all new links and
  we analyze quantitatively the tendencies used to create ties
  in these evolving online affiliation networks.  We first show that an
  accurate estimation of these probabilistic tendencies can only be
  achieved by following the time evolution of the network. For
  example, the actions that are attributed to the usual friend
  of a friend mechanism through a statistical analysis of
  a static snapshot of the network are overestimated by a factor of two.
  A detailed analysis of the dynamic network evolution shows that 
  half of those triangles were generated through other
  mechanisms, in spite of the characteristic static pattern.
  Inferences about the reason for the existence of
  links using statistical analysis of network snapshots must therefore
  be made with great caution.  Here, we start by characterizing every
  single link when the tie was established in the network.  This
  information allows us to describe the probabilistic tendencies of
  tie formation and extract sociological conclusions as follows.  The
  tendencies to add new links differ significantly from what we would
  expect if they would have not been affected by the individuals'
  structural position in the network, i.e., from random link
  formation.  We also find significant differences in behavioral traits
  in the social tendencies among individuals according to their degree
  of activity, gender, age, popularity and other attributes.
  For instance, in the particular datasets analyzed here,
  we find that women reciprocate connections three times as much
  as men and that this difference increases with age.
  Men tend to connect with the most popular people more often than women across all ages.
  On the other hand, triangular ties tendencies are similar, independent of gender,
  and show an increase with age.
  Our findings can be useful to build models of realistic social network
  structures and to discover the underlying laws that govern
  establishment of ties in evolving social networks.

\end{abstract}

\maketitle

\section{Introduction}

Uncovering patterns of human behavior addresses fundamental
questions about the structure of the society we live in. The choices
made at the individual level determine the emergent complex global
network underlying a given social structure \cite{Schelling}. Conversely, the
structure of the social network that constitutes an individual's
community also affects to a large extent the individual's ability to
act.
For instance, the position in the network structure may facilitate
one's ability to interact with others by providing
information of possible choices and their consequences
\cite{coleman}, or by supplying the individual with different
kinds of material and immaterial resources \cite{Bourdieu}. On the
other side, this structure may also
limit this individual's ability to act by excluding information
\cite{coleman} through local social norms and through social control.

\begin{figure}
\centerline{\includegraphics[width=.4\textwidth]{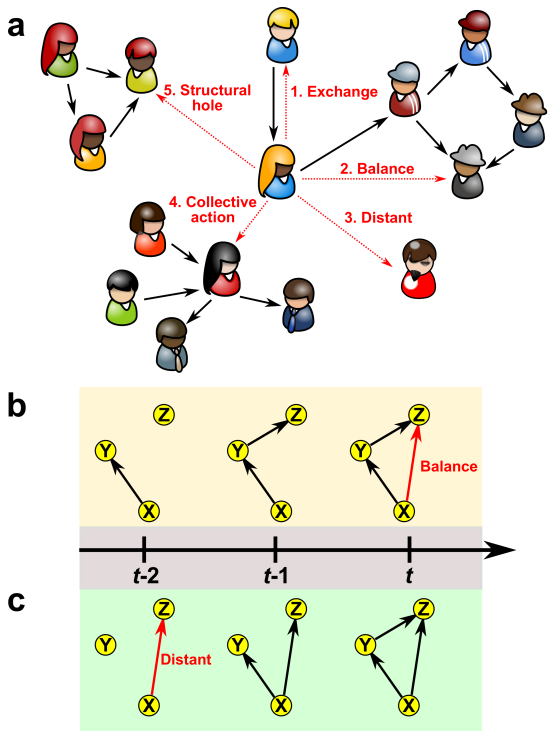}}
\caption{\label{example} (a) The five probabilistic
  tendencies we used to classify the interactions.  Black arrows
  indicate existing links and red arrows are the possible options for
  a new link, according to the following tendencies: 1. Social exchange,
  which corresponds to establishing a reciprocal
  link, i.e. add as favorite someone who has already added us to their
  favorite lists, 2. Balance, where we select a favorite who is in the
  list of one of our existing favorites (friend of a friend),
  3. Distant connection, where the connection is to a member with whom there is no
  proximity, i.e., one needs at least three links to reach this member,
  4. Collective action, where we connect to a person whose
  connectivity is well above the average connectivity in the
  community (we quantify this behavior by examining whether the total
  degree of the receiving agent belongs to the upper 5\% of the degree
  distribution at the given time), and 5. Structural hole, where a link
  connects two otherwise not connected clusters of at least 3 members each,
  and which are otherwise not directly linked to each other (in the picture this link
  would connect the cluster of people in hats with the red-haired cluster).
(b,c) Why we cannot extract tendencies from a static snapshot: in the presented example a
  triangle relation is built from time $t-2$ to time $t$ under two different
  scenarios that lead to the same resulting triangle.  
(b) The ties X-Y and Y-Z can be formed, at times $t-2$ and
$t-1$ respectively, via distant mechanisms resulting in a
balance mechanism for the formation of X-Z at time $t$. Here X uses a
friend of a friend to be introduced to Z. 
(c) A different path, though, would classify the
X-Z tie differently. If X connected to Z before connecting to Y,
then the X-Z link represents a
distant tendency, since there is no close connections between
them. A static network analysis would suggest that X used balance to connect to Z,
instead.
}
\end{figure}

Detecting regularities and motifs in the development of social
networks provides significant tools for the understanding of the
structure of society. Thus, a number of statistical association models
have been proposed to link a social network structure to a
statistically significant social mechanism of interaction \cite{Wasserman}.  Social
theoretical frameworks \cite{Monge}, like the MultiTheoretical
MultiLevel (MTML) formalism \cite{Contractor}, have proposed a set of
mechanisms of social interaction to describe the probabilistic
tendencies of creation, maintenance, dissolution, and reconstitution
of interpersonal ties during the evolution of a social network.
Examples of mechanisms include (see Fig~\ref{example}a): 1)
reciprocity (named {\it social exchange} after the most likely social
mechanism), 2) friend of a friend ties or closing triangles ({\it balance}),
3) exploration of distant network areas which require at
least 3 steps from the position of the person in the current network
({\it self-interest} theories), 4) ties facilitating dissemination of
information by linking to well-connected people (named {\it collective
  action} or {\it preferential attachment} \cite{barabasi}), and 5)
links that act as bridges between two sub-networks that are not
directly linked ({\it structural hole} mechanism).  Contractor {\it et
  al.}  \cite{Contractor} have further identified a set of
probabilistic tendencies for ties to be present or absent in networks
that the different families of theoretical mechanisms may cause.  One
important conclusion \cite{Contractor} is that a given family of
theoretical mechanisms may generate different probabilistic tendencies
for ties to be present or absent. Furthermore, the same probabilistic
tendency may be caused by several different families of theoretical
mechanisms.  In the present study we aim to unravel significant
patterns in these social mechanisms of human interaction by monitoring
and analyzing the time evolution of the actions of members of two online affiliation
networks. The term affiliation refers to data based on co-membership or co-participation in
events, where here members use the Internet to interact with each other through
the online sites \cite{borgatti}. A connection in such sites may indicate
underlying social ties \cite{davis}.

In principle, a formal statistical analysis, such as exponential
random graph models \cite{Wasserman,exprandgraph} would search for
regularities or motifs in the social structure by comparing a static
snapshot of the network with a suitable ensemble of equiprobable
random configurations.
However, this approach cannot characterize the decisions
taken (consciously or not) at the individual level on the type of
mechanism used for an established connection. A direct application of
a statistical analysis to evolving networks may not be able to resolve
the full spectrum of human interactions. This is due to the inherent
history-dependent nature of social interactions, i.e., the interaction
mechanisms determine the evolving network, which, in turn, conditions
the human choices of interaction. Figures \ref{example}b and
\ref{example}c illustrate this point during the generation of a
hypothetical triangular XYZ relation at time $t$. This static pattern
may be associated with a balance mechanism for the tie
XZ (friend of a friend) as a result of closing the triangle as shown
in Fig.~\ref{example}b.  However, a closer inspection of the time
evolution of tie formation reveals the possibility of a different
classification of the XZ link, where agent X has used the distant
mechanism at time $t-2$ to connect with Z as in Fig.~\ref{example}c.
As we show later, in Fig.~\ref{FIGchange}, the actual number of balance
links is over-estimated by a factor of 2 when we use static
snapshots of real communities.

The above example can be generalized to the global network level. For
instance, an agent may decide to connect to agents that are far away
in the network (distant mechanism).  Eventually, individuals are
brought closer to each other to form a tightly connected cluster.  The
evolving nature of the network may change those initial distant
interactions into balance, as new relations are created in the
network.  Therefore, the precise knowledge of the time evolution of
each tie in the network is crucial to unravel the relevant behavioral
mechanisms in a real community.

Here, we present a microscopic and temporal statistical analysis of
the evolution of two online social networks; one from its original
inception and the other after it is well developed. We aim to uncover
how the combination of different social mechanisms eventually shapes
the interaction network. Our longitudinal approach focuses on characterizing each
interpersonal tie at the time when it is established.  The knowledge
of the order in which each link was formed allows us to characterize
social patterns that cannot be derived from statistical analysis of
static snapshops of the networks.

\section{Datasets and methods}

We study the affiliation networks of
two online social networking sites in Sweden, {\it pussokram.com} \cite{pok_citation} and {\it qx.se} \cite{qx_citation}.
Both datasets were de-identified in their source.
The pussokram community (POK for
brevity) is used mainly by Swedish young adults for friendship,
including dating and non-romantic relations.  Activity in the
community was recorded for 512 consecutive days, starting on the day
that the site was created in 2001. At the end of recording, the
community had 28,876 members with a mean user age of 21 years who have
performed $\sim$190,000 interactions.
The QX site is the Nordic region's largest and most active web community
for homosexual, bisexual, transgender, and queer people. The site is also
frequently used by heterosexual men and women. Activity among the
users was recorded during two months starting November, 2005. At that
time there were $\sim180,000$ registered members; 80,426 of them were
active during the recording period establishing more than 1 million ties.

There are many types of interactions between members in the two
communities under study, but we focus on those which imply a firmer
commitment than, e.g., simply sending a message \cite{rybski}.  Such
interactions are (a) the favorites list in QX, and (b) the guestbook
signing in QX and POK.  The former interaction represents a clear
declaration of approval and/or interest, while the latter is a
communication publicly accessible to all community members where a
link does not necessarily indicate a particularly close relationship.
We compare two means of interaction in one community (favorites list
and guestbook signing in QX) and the same type of interaction
(guestbook signing) in two communities (QX and POK). We use the
guestbook signing to test consistent trends in the results.

In the QX dataset, it is possible that a user can remove a contact
at any point. There was a small number of such links, in total
less than 1\% of the total links, that were removed during our monitoring
window. It may be interesting to study the conditions
of ties removal in parallel with the addition process, but the small number
of removed contacts does not influence our results here, and we do not
further pursue this topic.

Each individual knows the following structural information from the
affiliation network: (a) who has added her in their favorites list or
who has written in her guestbook, (b) the members that she has added
in her favorites list and, (c) the friends of her friends since the
user can access the favorite list of friends.  This subnetwork defines
the immediate neighborhood of a member. Actions involving this
neighborhood are captured by social exchange and balance
mechanisms. The members situated farther away than this immediate
neighborhood are considered to belong to the rest of the network for
which the user has no direct information. Interactions with these
members are classified as distant.  A collective action can also be a
conscious choice, since a member can assess the popularity of others
through access to their favorite list, but it is also possible that this
action may not be conscious. Structural hole requires a much
wider knowledge of the network structure, and thus is the only
mechanism that a member does not realize that is using.

Our analysis can be readily extended to treat more general situations.
For simplicity, though, here we will not evaluate exogenous mechanisms
where interactions are based on attributes of the actors, such as
homophily, common interests, etc \cite{Contractor}.  We will further
not study the effect of focus constraints, i.e., the increased
likelihood of a tie being present among people that share a social
context, for example, living close to each other geographically or
working at the same office \cite{Feld}. The crux of the matter is to
quantify the different probabilistic tendencies about the actions of
the users as they are determined by the knowledge of the user about
the structure of the affiliation network that is the vital part of
his/her social life in the community.

The detailed quality of our longitudinal data allows us to identify the precise
probabilistic tendencies for tie formation that a newly established
link corresponds to, when an actor adds a new favorite in his list (or
signs a guestbook). Every interaction that occurred between two
members was recorded together with the timestamp when the event took
place. We create the evolving network of interacting agents by adding the
directed links in sequential order. For example, at the time when a
member X adds a member Y in the favorite list of X, we create a
directional link from X to Y.  Similarly, in guestbook signing, the
directional link from X to Y corresponds to X writing in Y's
guestbook (we take into account only the first time X signs Y's guestbook
and ignore repeated signings).
Every time we add a link, we characterize this action
according to the probabilistic tendencies described in
Fig.~\ref{example}a, as dictated by the network configuration at the
given moment. Every link is therefore assigned to one or more
probabilistic tendencies: exchange, balance, distant, collective
action, and structural hole. We define the probabilities of each
tendency $P_{\rm exc}$, $P_{\rm bal}$, $P_{\rm dis}$, $P_{\rm ca}$,
and $P_{\rm sh}$ respectively, as the number of links that were
created using the corresponding tendency normalized by the total
number of links created up to a given time $t$.

A newly formed link is assigned to the exchange tendency when it is
established in the opposite direction of an existing link. The balance
tendency corresponds to a directed network distance $\ell= 2$, i.e.
when a link points to a friend of a friend ($\ell$ is the directed
distance between two nodes just before the link is formed - defined as
the shortest path with all arrows pointing to the same direction, so
that a directed path exists between these two nodes). If the distance
between the two nodes is $\ell \ge 3$, the link represents the distant
tendency. A link is considered as collective action when the chosen
node is a hub. We define a hub as a node whose total degree (counting
both incoming and outgoing links) belongs to the upper 5\% of the
degree distribution as measured at the time of link formation. A link
represents the structural hole tendency when this link connects two
clusters of at least three members each that would otherwise be
disconnected.  Table 1 summarizes these definitions.

\begin{table}[ht]
\caption{List of tendencies, indicators, and the type of 
  directionality in the network used to detect the tendency. 
  $\ell$ is the distance between two nodes as measured by the shortest path 
  in the directed network.
}
\begin{center}
\begin{tabular}{|l|l|l|l|}
  \hline
  & Tendency	& Indicator	& Directionality\\
  \hline
  $P_{\rm exc}$	& Social Exchange	& $\ell=1$, mutual link	& directed\\
  $P_{\rm bal}$	& Balance	& $\ell=2$ 	& directed \\
  $P_{\rm dis}$   & Distant		& $\ell\ge 3$	& directed \\
  $P_{\rm ca}$ 	& Collective Action 	& link to a hub	& undirected\\
  $P_{\rm sh}$ 	& Structural Hole	& connect two clusters	& undirected\\
  \hline
\end{tabular}
\end{center}
\label{table}
\end{table}

In general, the increase in the probability of a tie forming under
a given tendency will not necessarily be compensated for
by a tie with decreased probability under another tendency.
The relative probabilities between tendencies do not necessarily
present competing risks and different tendencies may act at
the same time.
It is then possible that one link jointly represents more
than one type of tendency in tie formation. In this case, we assign
this action to all involved tendencies.  For instance, a balance tie
could be also catalogued as collective action if the agent closes a
triangle by connecting to a hub. Based on the definitions, only
balance and distant tendencies are complementary to each other
($P_{\rm bal}+P_{\rm dis}=1$) so that the presence of one excludes the
presence of the other. The other tendencies are normalized as, e.g.,
$P_{\rm ca}+P_{\rm not-ca}=1$ ($P_{\rm not-ca}$ is the probability of
not performing a collective action). 

By establishing all links in the order they appeared, we can recreate
the entire history of the directed network of interactions. While POK
starts at $t=0$ from an empty network, QX has a large part of the
network already in place at $t=t_0$, our initial recording date. In this case, we know all the
existing links at $t=t_0$. Thus, in QX, we characterize only the network
links that were added during the monitoring period.

\begin{figure}
\centerline{\includegraphics[width=.5\textwidth]{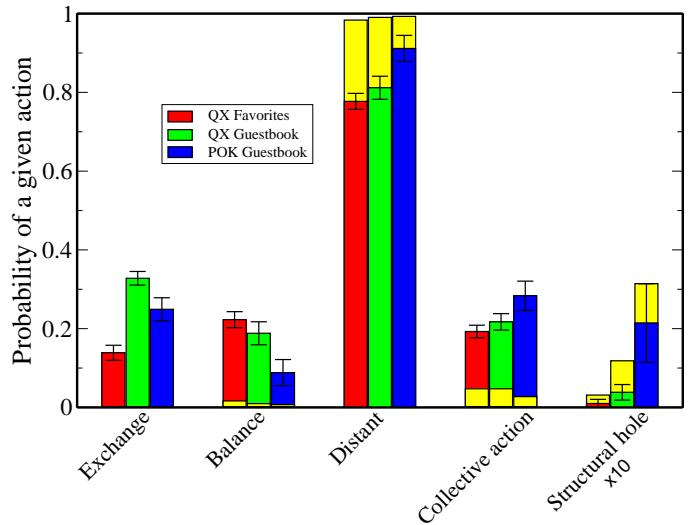}}
\caption{\label{FIGtotal_average} The relative appearance of the five
  probabilistic tendencies in the actions of the community members in
  QX using favorites (red), in QX using guestbook (green), and in POK
  using guestbook (blue). These tendencies are compared to a
  completely random selection (yellow).  Exchange and balance are
  practically non-existent in random selections, but carry significant
  weight in the interactions of the real communities. Connecting to
  distant members appears in the community much less frequently than
  in random, while the preference towards well-connected agents
  (collective action) is significantly more prominent. Finally,
  structural hole is significantly suppressed in the real communities
  compared to the randomized case.}
\end{figure}

Figure ~\ref{FIGtotal_average} presents the fraction of appearance of
each tendency when considering all recorded interactions in the
studied datasets, QX and POK, and the means of interaction, guestbook
and favorite list.

The results are fairly independent of the specific community and the
means of interaction. The probabilities $P_{\rm exc}$, $P_{\rm bal}$,
and $P_{\rm ca}$, appear each at approximately 15-30\% of all actions.
The distant mechanism is dominant, with $P_{\rm dis}\approx 80$\% of
the established links.  Collective action remains low at $P_{\rm
  ca}\approx 20$\% considering that this tendency is considered the
main driver in some models of network formation through preferential
attachment \cite{DeBlasio,barabasi}.  A very small fraction of links
$P_{\rm sh}$ `fills' the structural holes.  This is a result of the
small numbers of clusters that exist in each community, so that the
chances to connect isolated clusters are small.  In particular,
comparison to the random case (where the same members act at each time step,
but instead of the established link they choose a random connection,
Fig.~\ref{FIGtotal_average}, yellow bars) reveals that the structural
hole tendency is more probable when an agent connects to a random member.
In other words, although there exist opportunities for structural
hole, the members tend to stay within their own sub-networks, despite
the lack of knowledge on the global structure. The percentages for the
other tendencies are also very different from random selections. This
implies that community members follow social criteria when adding new
favorite members (or sign guestbooks). We verified the robustness
of our results by comparing the percentages of the links at the early
stages of network formation with those of the links that were established
later in the process. For example, in QX favorites the first half
of the actions dataset gives practically the same result as the second half:
exchange was 13.8\% for the first half and 13.9\% for
the second, balance was used 22.1\% versus 22.4\%, and collective action
was used 18.8\% versus 19.7\%. Furthermore, the stability of this result over the
evolution of the links is verified later, in Fig.~\ref{FIGchange}.

Our analysis has shown that the direct calculation of the tendencies
of link formation from the time evolution of the network provides a
consistent characterization of the social mechanisms involved, which
is different from a static snapshot.  Furthermore, the present
analysis allows to determine if the found tendencies are influenced by
important actor attributes that are hypothesized to have an
association with ties formation \cite{parker-asher}. These attributes
include age, gender, popularity and activity intensity measured as the
number of links developed at a given time. Next, we incorporate these
attributes in our analysis to attempt to understand how different factors
influence the behavior of the actors. We show that the gender, age,
activity intensity, and popularity can lead to a different probability
of using a given tendency.

\section{Results}

\subsection{Gender influence}

\begin{figure}
\centerline{\includegraphics[width=.48\textwidth]{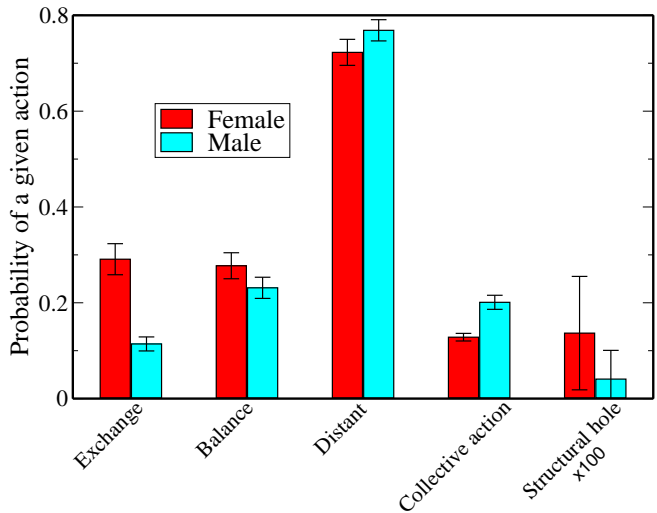}}
\caption{\label{FIGgender} Probability of different tendencies, based
  on self-reported gender in the QX community in favorite list
  interactions.  Exchange and structural hole are significantly more frequent in females
  compared to males.}
\end{figure}

Our analysis reveals that gender is an important attribute determining
the social tendencies. Analysis of the QX community (the only one
reporting gender) reveals that men do not use some mechanisms in the
same way as women (Fig.~\ref{FIGgender}). Using the gender information
in the QX favorite lists, we find that a female member is almost three
times more probable to have an exchange tendency compared to male
members and three times more probable to fill structural holes
(men, on the other side, perform distant and collective actions at higher
percentages). The significant difference in exchange, for example,
reveals a different approach of online
communication between men and women \cite{Herring}.  Our result is in
agreement with the self-reported tendency of women users to exchange
more private e-mails than participating in public discussions
\cite{Hoffman}. The stronger preference for exchange of female users
in the QX community can also be seen as a similar trait where women
tend to develop stronger inter-personal relations by frequently
reciprocating friendships.

\subsection{Age influence}

\begin{figure}
\centerline{\includegraphics[width=.48\textwidth]{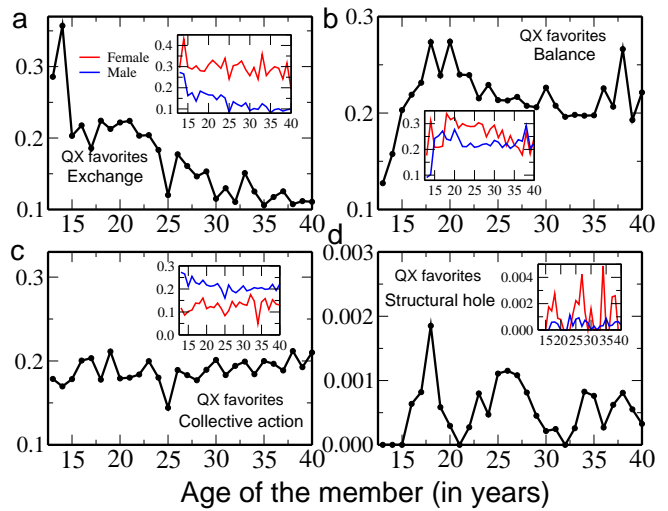}}
\caption{\label{FIGage} Variation of the average tendency percentage
  with the self-reported age of the QX members. (a) The exchange tendency
  decreases with age. (b) The balance tendency
  is sharply increasing with age in younger ages, and slowly
  declines for ages above 20. We do not observe any strong dependence on age
  for (c) collective action or (d) structural hole (bottom right).
  The insets show the
  differences between males and females of the same age for each
  tendency.  }
\end{figure}

In the databases that we studied, members of different age tend to
present different behaviors.  In Fig.~\ref{FIGage} we calculate the
fraction of actions that correspond to a tendency as a function of the
self-reported age of the QX members. In the insets, we separate the
corresponding probabilities for male and female members.  We find that
while reciprocity in women remains high as they age,
men instead reduce it by a factor of 2 as they reach 40.
This shows that younger male members are more eager to reciprocate their
connections.
In contrast, the level of balance is
roughly constant for both genders and independently of age,
with an important exception at the youngest ages,
where members younger than 20 years old are using systematically less
balance links. This could be because it is more difficult for
them to develop a stable local network in an adult-oriented community.
There are no significant trends with age for collective action or
structural hole, although the latter tendency is rarely used.
The gender-based trends shown in Fig.~\ref{FIGgender} are consistent
with the age-based results. Women of a given age are always using more
exchange and less collective action tendencies than men of the same
age (insets of Fig.~\ref{FIGage}).

\subsection{Activity influence}

Communities include members of varying activity \cite{rybski}.  In
order to study the effect of the different activity levels, we address
the question of whether a higher involvement in a community is
accompanied by a different pattern in the probabilistic tendencies of
social mechanisms. We calculate the different probabilities of social
mechanisms as a function of the number of $k_{\rm out}$ outgoing links
for each member. For instance, $P_{\rm \alpha}(k_{\rm out})$ (where
$\alpha$ denotes exchange, balance, etc) measures the probability that
the next action will correspond to $\alpha$, when the member has $k_{\rm out}$
outgoing links.
We measure 
$P_{\rm \alpha}(k_{\rm out})$ through all the actions of members when
they increase the number of outgoing links from $k_{\rm out}$ to
$k_{\rm out}+1$, irrespectively of the time that the action was
performed. Interestingly, we find that a member typically modifies
his/her behavior according to its current degree of activity $k_{\rm
  out}$.  As a member becomes more involved in the community and, as a
consequence, increases the size of his/her favorites list or signs
more guestbooks, the member switches to a different relative
percentage of using each tendency.

\begin{figure}
\centerline{\includegraphics[width=.45\textwidth]{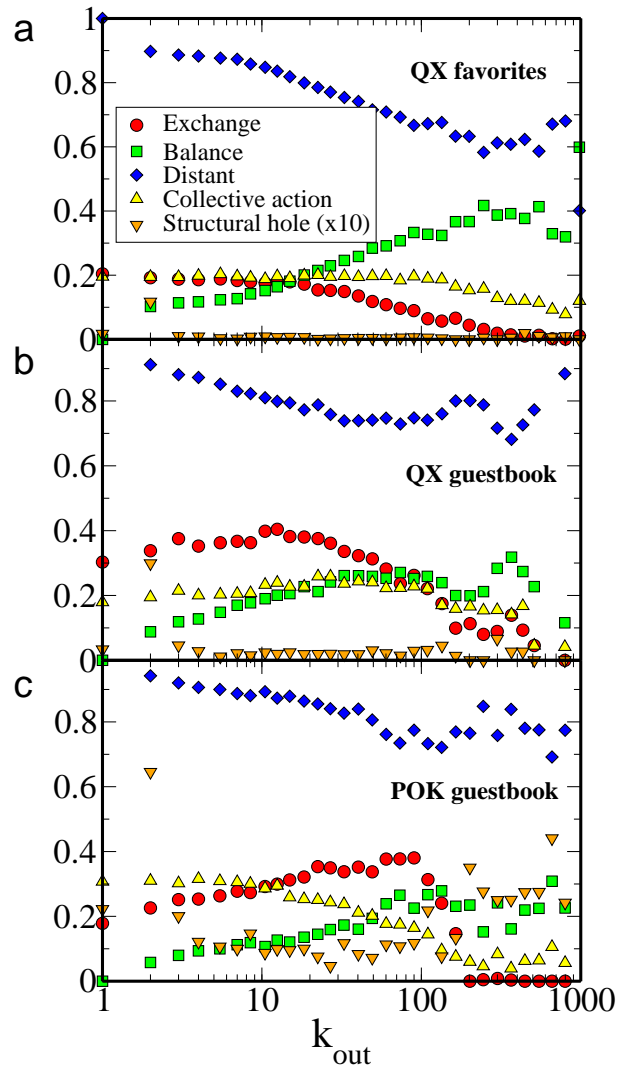}}
\caption{\label{FIGevolution} Fraction of the appearance of a tendency
as a function of the adding member's list size, at the time of addition.
Qualitatively, all three datasets are in agreement with each other.
The small quantitative differences may be due to the different means
of interaction and/or the design of each platform.
}
\end{figure}

We identify the following pattern which is very consistent across the
two datasets and different types of interactions (see
Fig.~\ref{FIGevolution}). The first tie of a new member is always
distant since the member has no network established.  However,
even at this stage, 20-30\% of these links are also exchange---
meaning that a new member readily `responds' to the incoming link by
established members--- and collective action, meaning that the member
immediately searches for popular members in the community. At this
earlier stage, balance tendency is suppressed, since linking to friends
of friends requires first a firm establishment of the immediate
neighborhood.

An interesting crossover appears when the members arrive to a size
$k_{\rm out}\approx 10$ in their favorites list (see for example
Fig.~\ref{FIGevolution}a for QX favorites).
The percentage of all tendencies up to that value is approximately
constant. At around 10 interactions in QX favorites, balance overtakes
both exchange and collective action in the behavioral tendencies.  As the
members keep adding more links, the distant mechanism drops
significantly to approximately 60\% after $k_{\rm out}\approx 100$,
and the balance tendency grows increasingly stronger consequently.
Similarly, the exchange tendency declines steadily towards 0 as the
size of the favorites list increases towards the hundreds.  Collective
action leading to preferential attachment seems to be the most stable
over a longer $k_{\rm out}$-range. Finally, the relative probability
of $P_{\rm sh}(k_{\rm out})$ peaks at low and large values of $k_{\rm
  out}$. The structural holes are filled mainly by either new members
 or well-established members, with a significantly smaller fraction of
 structural holes performed in the intermediate $k_{\rm out}$
 regime. This interesting behavior reveals trends in the social
tendencies across the individual users as they enter the network.

The choice of different tendencies is, thus, shown to have a complex
dependence on the individual's level of activity.  In addition to
external attributes, such as gender and age, we find that very
active members have different tendencies than the less active
ones. Such features can only be extracted by following the entire time
evolution of each member's connections.

\subsection{Popularity attributes}

So far, our analysis focused on quantifying the different
probabilistic tendencies as seen from the member that establishes a
link. We characterized the outgoing links which can be controlled by
their initiator, in the sense that any member can choose where, when,
and how often connects to other members. However, the popularity (or
attractiveness) of a member cannot be adjusted at will. We
characterize the popularity based on the number of incoming
links. Using the same methodology as above, we can now study how
different tendencies determine the popularity of a member.

For each relationship between two people we assign the initiator,
i.e. the member who contacted the other member first, and the
receiver, i.e. the member who was contacted. In the case of a
reciprocal relation we only characterize the link that was established
first.  Given the list of a member's connections, we can then know
what fraction of those connections is due to the initiative of this
member and what fraction originated from the other side.  Thus, if
someone very often reciprocates but seldom initiates links, she will
have a small value of initiated links although she may have a large
number of incoming and/or outgoing links.

\begin{figure}
\centerline{\includegraphics[width=.24\textwidth]{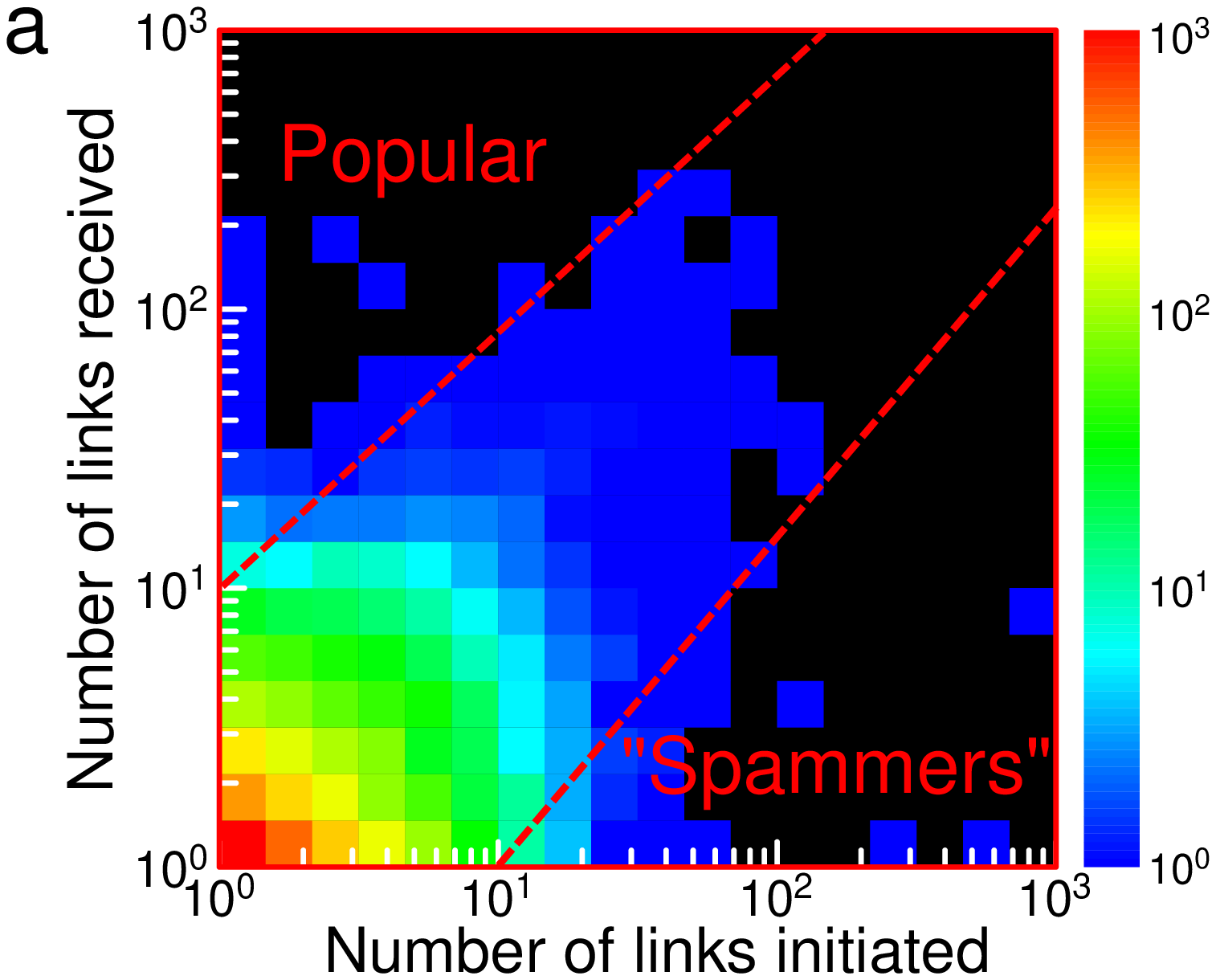}
\includegraphics[width=.24\textwidth]{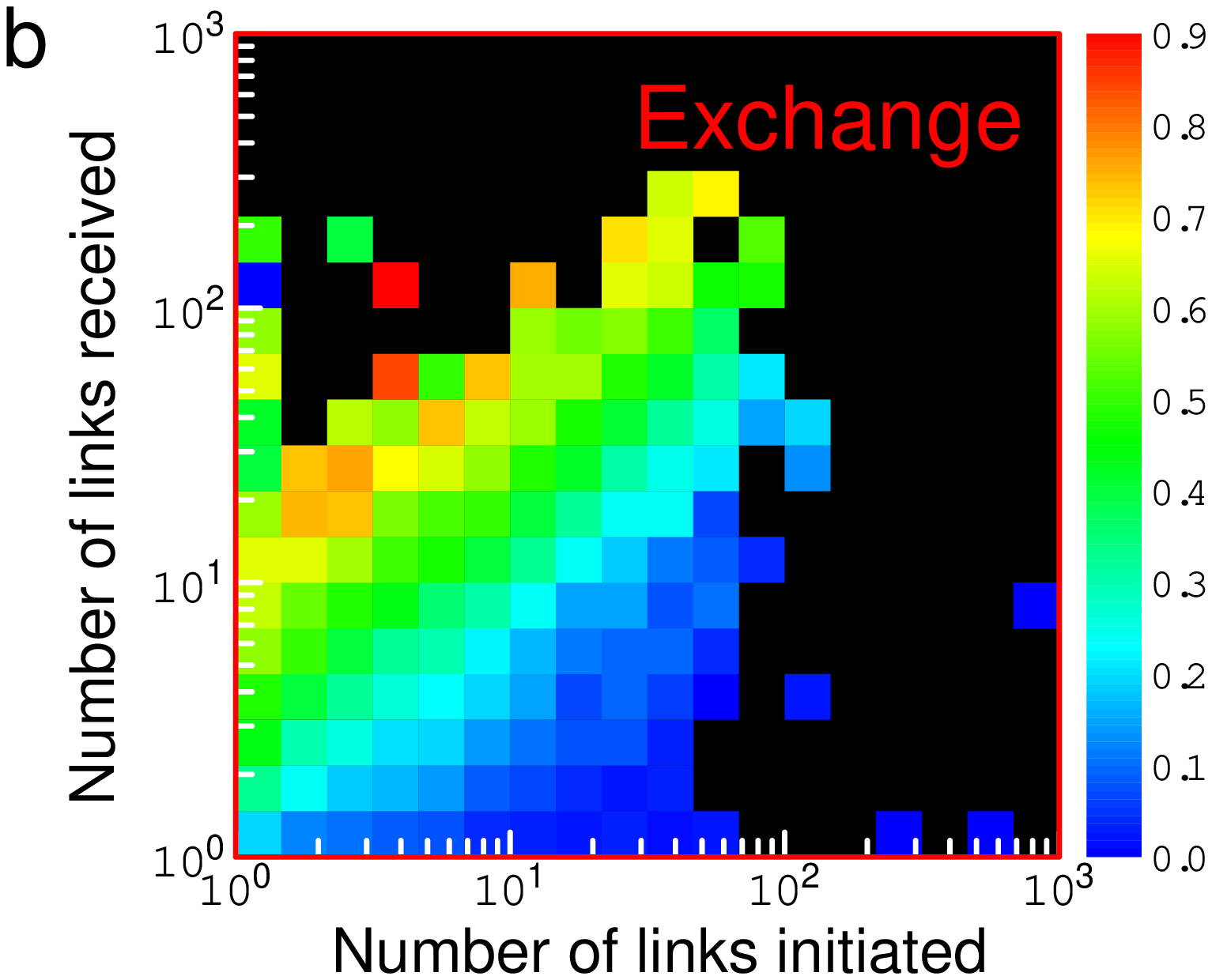} }
\centerline{
\includegraphics[width=.24\textwidth]{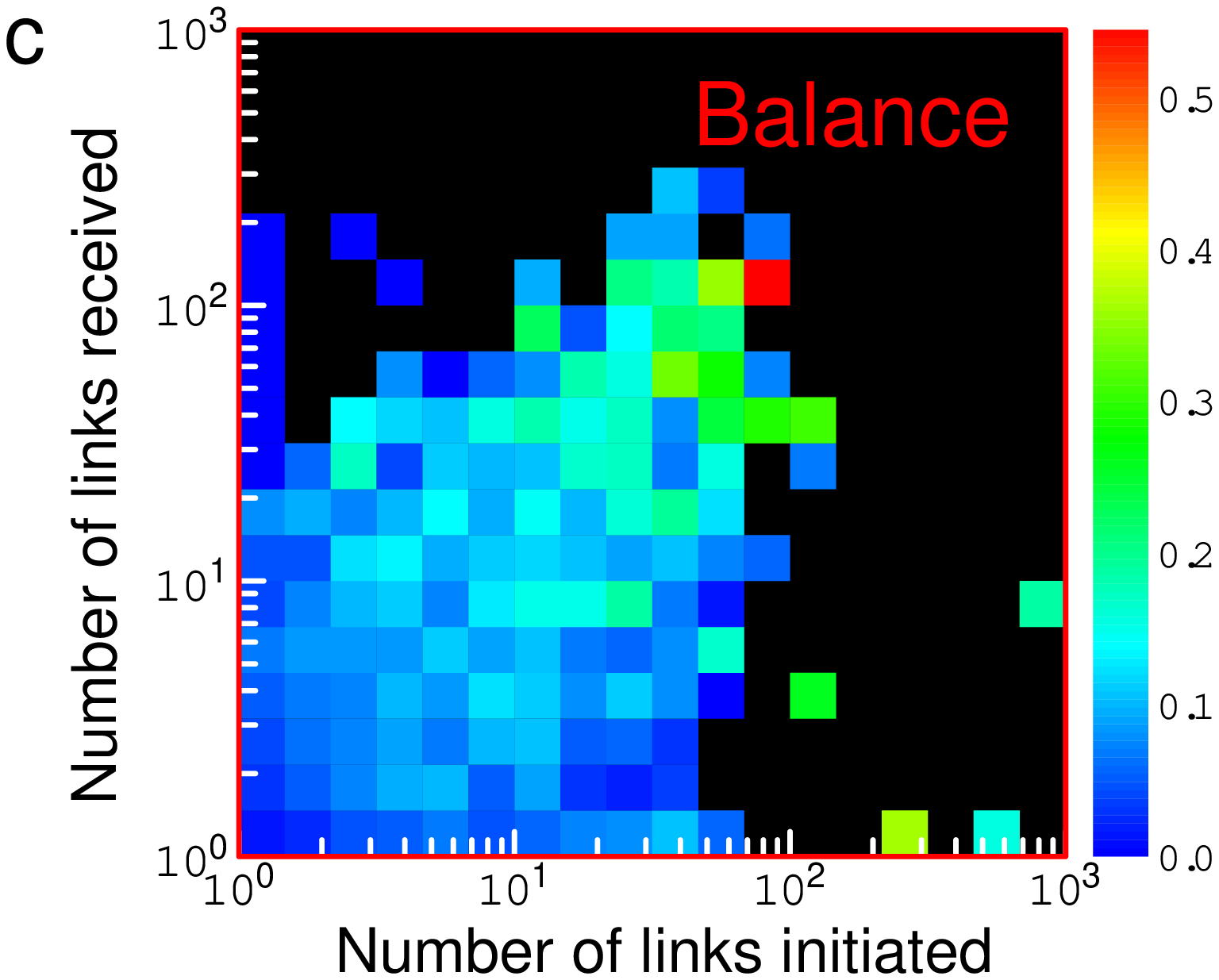}
\includegraphics[width=.24\textwidth]{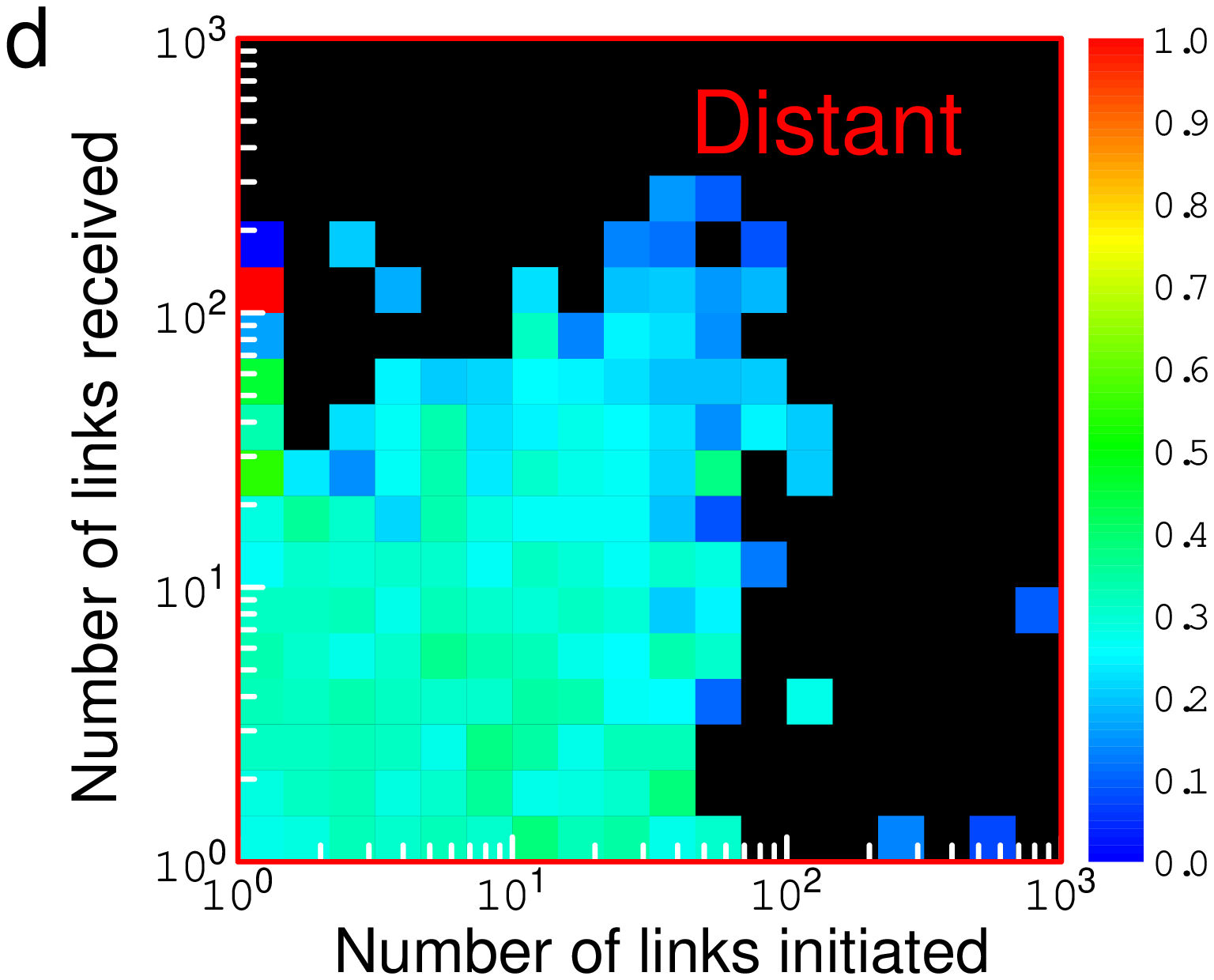} }
\caption{\label{FIGpopular} (a) Histogram of the number of members as
  a function of the links that they initiated (x-axis) and the links
  that were pointed to them but initiated at the partner's side
  (y-axis). (b-d) Average percentage of exchange, balance, and distant mechanisms
  as a function of the links initiated and received.
}
\end{figure}

In Fig.~\ref{FIGpopular}a we present the histogram of how many members
fall into each category. The diagram is roughly divided into three
areas: a) Members who initiate a lot of connections but are first
contacted by very few members (`spammers') b) Members who on average
equally initiate and receive contacts, and c) Members who receive many
more contacts than they initiate (`popular').

The importance of using the time evolution of probabilistic tendencies
to determine behavior is reflected in this popularity
classification. In Figs.~\ref{FIGpopular}b-d
we present the average percentage for each
category and for each tendency that the members use when they add
friends themselves. The exchange tendency shows a clear variation with
respect to this classification. The `popular' members in the upper
diagonal part of the distribution use a lot of exchange, which can be
understood since they respond to friendship requests but rarely start
new connections. As we move towards the `spammers' the exchange
tendencies almost disappear, since very few people approach those
members and therefore they have small chance to use exchange. On the
contrary, the spammers tend to use balance more, i.e. they connect to
friends of friends, since they try to access the largest possible
number of the accessible members (Fig.~\ref{FIGpopular}c).
Finally, connecting to distant parts of the network (Fig.~\ref{FIGpopular}d)
has a more uniform behavior, although the popular members seem
to use it more, pointing to a ``rich-club'' phenomenon \cite{richclub}.

The above described trends demonstrate the richness of information
that becomes accessible by following the evolution of link formation.
Nevertheless, we next show that even in the absence of the network history,
we can still deduce some useful conclusions on the probabilistic
tendencies.

\begin{figure}
\centerline{\includegraphics[width=.48\textwidth]{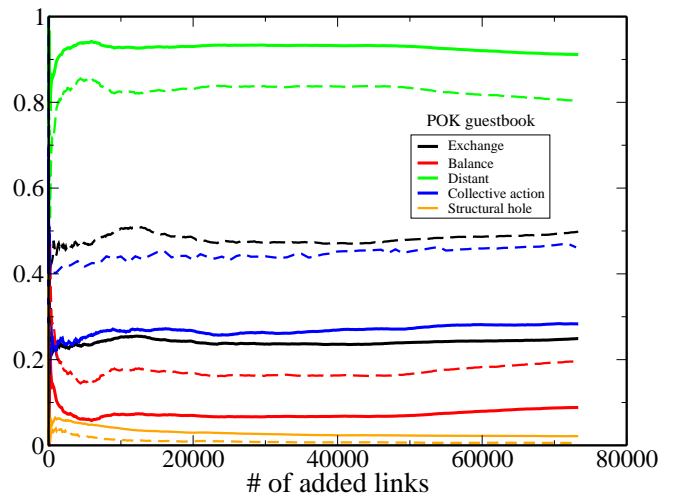}}
\caption{\label{FIGchange} Comparison of the probabilistic
  tendencies fraction, where links are characterized either at the
  time of addition (solid lines) or at the time of observation (dashed
  lines) in the POK community.
}
\end{figure}

\subsection{Neighborhood landscape change}

As discussed above, the presented analysis would not be possible
without continuously monitoring the time evolution of the links.  The
characteristics of a given link with time do not remain necessarily
the same as when the connection was established, but they can change
due to the addition of more links or the removal of existing ones.
For example, a friendship that starts between two isolated individuals
may evolve into a densely connected neighborhood, so that a link that
started as distant may eventually switch with time to either balance,
exchange, collective action, structural hole, or any combination of
them.

In order to study how significant the evolution of the link formation
tendencies is, we compare the probabilistic tendencies obtained above
following the time evolution with those obtained by a statistical
analysis of a snapshot of the network.  The statistical analysis of
the static snapshot is done by characterizing all existing links at the
given time without using the information from the time when the link
was established.  Thus, we remove a link and characterize it as if it
was just established, and immediately re-insert it back in the
network. Thus, each link is assigned to the specific probabilistic
tendencies according to the current neighborhood environment of each
agent, independently of the time it was established. We repeat this
process for all links in the static snapshot and we calculate the
relative percentage for each mechanism.

In Fig.~\ref{FIGchange} we compare the running percentages for each
tendency at the moment of addition, such as those measured in
Fig.~\ref{FIGtotal_average}, to those of the corresponding static
network. All tendencies are different in these two measurements.
Exchange is the only predictable tendency, since by definition it
appears two times more at the time of observation compared to the time
of addition.  The other tendencies cannot be predicted from the static
measurements.  For example, although a member is typically using the
balance tendency to add links at a percentage of around 10\%, if she
tries to evaluate her neighborhood at any point in time she will find
out that now approximately 20\% of her acquaintances fall under the
balance theory. Similarly, the central hubs seem to be re-enforced,
since collective action is used in less than 30\% of the total
actions, but eventually more than 45\% of the links are directed
towards the biggest hubs. In other words, members are ultimately
attached to hubs more often than we could conclude from characterizing
their original actions only, due to the dynamic
environment. This quantifies and generalizes the situation depicted
in Figs. \ref{example}b and \ref{example}c: the knowledge of the network structure
at a given time is not sufficient for characterizing the probabilistic
tendencies.

Another aspect of this plot (Fig.~\ref{FIGchange}) is that the
tendencies at the time of addition reach their asymptotic values quite
fast and they remain roughly constant with time. The corresponding
values extracted from the static networks are also quite robust and
follow closely the variations of the values in the evolving networks,
creating a constant gap between the two curves. Since there is
currently no method to estimate the magnitude of the difference
between the two cases by static information only, it is still not
possible to extract the percentage of the probabilistic tendencies
without following the network evolution.

Next, we compare our results with other directed social interaction networks
from the literature, such as the Epinions \cite{epinions}, SlashDot \cite{slashdot}
and LiveJournal \cite{livejournal} communities.
The datasets were downloaded from http://snap.stanford.edu/data.
The Epinions dataset is a directed network of trust from epinions.com, where a user
can declare her trust towards another user, based on submitted reviews. This trust creates a directed link between the two
users. The network has 75879 nodes and 508837 links.
Slashdot.com is a technology-oriented news site, where users can tag
each other as friends or foes. In our analysis we only use the friendship links. We use two snapshots
of the network, on November 6, 2008 (77360 nodes and 905468 links) and on February 2, 2009 (82168 nodes
and 948464 links) \cite{slashdot}. Finally, Livejournal.com is a social networking site, where users can declare
who they consider as their friends. The network that we use has 4847571 nodes and 68993773 links.
For these networks we only
have the static snapshots. Therefore, we can only study the exchange
tendency, which is the only one that remains unmodified in a static
network (we can always measure the existence of reciprocity,
independently of the time it was established).

\begin{figure}
\centerline{\includegraphics[width=.45\textwidth]{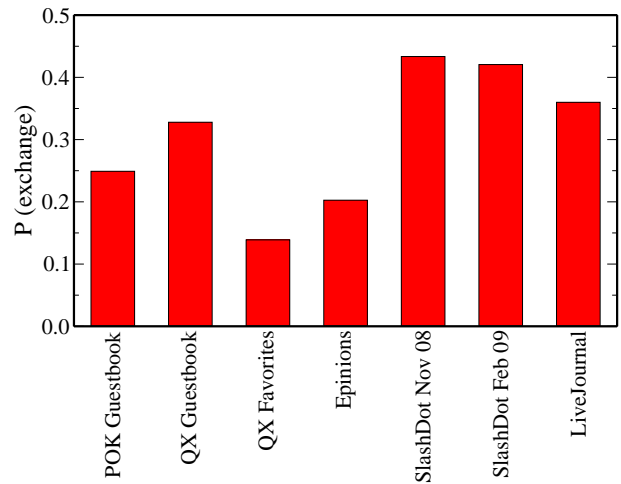}}
\caption{\label{FIGmore} Probabilistic exchange tendencies extracted from
  static network snapshots for several directed networks.  }
\end{figure}

The probability to use the exchange tendency among the different social
networks (Fig.~\ref{FIGmore}) depends on the specific features of
each community.  For example, in the SlashDot and in the LiveJournal
communities, where a link shows that a user declares another user as
being his/her friend, there is a large degree of the exchange tendency
because mutual relations are favored in these social networking
environments. In contrast, in the QX database the exchange tendency is
quite smaller due to the nature of this community. Similarly, in the
Epinions database a link shows that a member trusts the tech reviews
of the other member, but this relation is usually not mutual (if I
trust the reviews of an expert reviewer, this reviewer may not
necessarily trust my reviews).

\section{Discussion}

The wealth of information obtained by our longitudinal analysis can complement
other statistical analysis for probabilistic tendencies
\cite{Contractor,exprandgraph}. The family of exponential
random graph models \cite{Frank} ($p^*$), and in particular the logit $p^*$ models \cite{Wasserman},
have been very successful in analyzing network snapshots
at a given moment in time. These methods
detect network patterns that appear more frequently than a random null
hypothesis would assume. In this way, the underlying
mechanisms of network creation are inferred from the resulting motifs. 
Our present analysis goes beyond this approach by directly facing
a number of key issues: we can follow the entire network evolution,
we can characterize individual actions, and we can also assign known
mechanisms to any given action. The results of
these actions often yield network patterns where an individual
contribution may be lost in the static snapshot pattern, due to the effect
of subsequent connections. In broad terms, our analysis compared to exponential
random graph models may be considered to be the analogue of
a microscopic statistical physics description compared to a macroscopic
thermodynamic approach.

Here, we have shown that following the order of links establishment at the
microscopic level in a social network provides a direct measurement of
the probabilistic tendencies. This allows both the quantification of
the relative strength between tendencies in a given community, and the
extraction of useful sociological conclusions. For example, in the
communities that we studied, we show that women tend to use the
exchange mechanism more frequently than men. This tendency is
more pronounced with age since reciprocity in older men largely declines
while in women it remains stable across all ages.
In these communities, also, men tend to connect to the hubs
more often than women, independently of age. The use of triadic closures
is almost constant for both genders and all ages,
except for the youngest members with ages below 20. This may be
a consequence of the more adult-oriented character of the community.
Similarly, we capture a different use of the tendencies
between the more active and less active members. The basis
of our findings is that these results cannot be derived analyzing a
snapshot of a static network.  As shown in Figs. \ref{example}b and \ref{example}c
and quantified in the preceding section, it is not possible to make
assumptions of why a link exists a long time after the link was
established.

Our findings reflect the behavior of users in the online networking
sites that we studied. The suggested method of following the dynamic
evolution, though, represents a consistent method which can be applied
to other networks. Further studies in different online communities should
elaborate on whether the trends reported here with respect to sex, age,
etc, are generic to other types of networks.

The present analysis complements other approaches in the literature \cite{leskovec}
by focusing on individual actions and the study of how the underlying
mechanisms behind these actions are driving the evolution of the
large-scale social network. The ability to isolate individual actions
can be also very useful in studying behaviors that are unusual, and
help characterize idiosyncratic ways of building the friendship
network.  The present analysis can be extended to exogenous
mechanisms, as well, by incorporating information from other aspects
of the activity in the community (e.g. joining specific clubs,
participating in forum discussions, communities, etc).

\begin{acknowledgments}
The study of the de-identified QX site network data was approved by the
Regional Ethical Review Board in Stockholm, record 2005/5:3.
We are thankful to Brian Uzzi and Hern\'an D. Rozenfeld for valuable discussions.
We acknowledge support from NSF and the ARL
under Cooperative Agreement Number W911NF-09-2-0053.
SH thanks the ONR, DTRA, DFG, EU project Epiwork,
and the Israel Science Foundation for financial support.
FL acknowledges Riksbankens Jubileumsfond for financial support.
\end{acknowledgments}

\end{document}